\documentclass[final, numberedheadings]
  {aipproc}

\layoutstyle{8x11single}

\begin{document}

\setcounter{bottomnumber}{2}
\renewcommand{\bottomfraction}{0.95}

%\title{A non-linear resonance model for the black-hole and neutron-star QPOs
% - some links between observation and theory}
 
%\title{Orbital resonance model for black-hole and neutron-star QPOs: links between observation and theory}
\title{A non-linear resonance model for the black hole and neutron~star
QPOs: theory supported by observations}


\author{Gabriel T\"or\"ok}{
address={Institute of  Physics, Faculty of Philosophy and Science, Silesian University in Opava, Bezru{\v c}ovo n\'am. 13, CZ-746 01 Opava, Czech Republic}
}

\author{Marek Abramowicz}{
address={Department of Physics, G{\"{o}}teborg University, S-412-96 G\"oteborg, Sweden;~~also~$^\ast$ and $^\ddagger$}
}

\author{W{\l}odek Klu\'zniak}{
  address={Institute of Astronomy, Zielona G\'ora University
        ul. Lubuska 2, PL-65-265 Zielona G\'ora, Poland},
       altaddress={Copernicus Astronomical Center, ul. Bartycka 18, 00-716 Warszawa, Poland}
}

\author{Zden{\v e}k Stuchl\'{\i}k}{
  address={Institute of  Physics, Faculty of Philosophy and Science, Silesian University in Opava,
        Bezru{\v c}ovo n\'am. 13, CZ-746 01 Opava,
	Czech Republic}
}

\begin{abstract}

Kilohertz Quasi-Periodic Oscillations (QPOs) have been detected
 in many accreting  X-ray binaries.
 It has been suggested that the highest QPO frequencies
 observed in the modulation of the X-ray flux reflect a non-linear
 resonance between two modes of accreting disk oscillation.
 This hypothesis implies certain very {\it general predictions},
 several of which have been borne out by observations. Some of these
 follow from properties of non-linear oscillators, while the
 others are specific to oscillations of fluid in strong
 gravity. A 3:2 resonant ratio of frequencies can be clearly recognized
 in the black-hole as well as in the neutron-star QPO data.
 
\end{abstract}

\maketitle

%%%%%%%%%%%%%%%%%%%%%%%%%%%%%%%%%%%%%%%%%%%%
%% MAINMATTER
%%%%%%%%%%%%%%%%%%%%%%%%%%%%%%%%%%%%%%%%%%%%
%\end{document}
\section{Introduction}

The resonance model of \citet{KluAbr:2000:UNP:} explains pairs of high
frequency QPOs observed in neutron star
and black hole (microquasar) sources in Galactic low mass X-ray binaries
as being caused by a 3:2 non-linear resonance between two global modes of
oscillations in accretion flow in strong gravity
\footnote{For a recent review of observations see \cite{Kli:2005:ASN:} and for a review of the
theory see \cite{Abr:2005:ASN:}.}.
 In this report, we shortly review rather convincing observational arguments that support the
resonance model:

\begin{enumerate}
\item {\it The rational frequency ratio 3:2}.~ For microquasars
, the observed ratio $\nu_\mathrm{u}/\nu_\mathrm{l}$ of the upper and lower frequency is
$3/2$. For neutron star sources the ratio varies in time, but its
statistical distribution peaks up, within a few percent, at the $3/2$
value.   

\item  {\it The inverse mass scaling}.~ In compact sources, $\nu_\mathrm{u} \sim 1/M$.

\item {\it The frequency-frequency correlation}.~ The
upper and lower frequencies of a particular neutron star source are linearly
correlated along the \emph{``Bursa line''}, $\nu_\mathrm{u} = A\,\nu_\mathrm{l} + B$, with $A \not
= 3/2$. Observational points occupy a finite sector of the Bursa line
which typically crosses the reference line $\nu_\mathrm{u} = (3/2)\,\nu_\mathrm{l}$ at the
\emph{``resonance point''}.
\footnote{Only a few examples of sources for which the actually known data do not
  cross the 3:2 line are known \citep[see, e.g.,][]{Lin-etal:2005:APJ:}.}

\item  {\it The slope-shift anticorrelation}.~ For a sample of
several neutron star sources, the coefficients $A\,,B$ of the Bursa lines
corresponding to individual sources in the sample are anticorrelated, $A
= 3/2 - B/(600\,{\rm Hz} \pm \Delta \nu)$, where $\Delta \nu \ll
600\,{\rm Hz}$ is a small scatter. 

\item  {\it Behavior of the rms amplitudes across the resonance point}. For the few sources examined so far, the difference between rms amplitudes of the lower and upper QPO changes sign at the resonance point.
 
\end{enumerate}

\par One should have in mind that except the inverse mass scaling \emph{the above properties are typical of a
non-linear resonance in any system of two weakly coupled
oscillatory modes. They give no direct clues on how to answer several
fundamental questions connected to the specific physics of the
oscillations of accretion disks:}

\begin{itemize}

\item {\it Which are the two modes in resonance?} An often discussed
possibility is that they are the epicyclic modes \citep[see, e.g.,][]{AbrKlu:2004:AIP:}, but there are some
difficulties with this interpretation.

\item {\it What are the eigenfrequencies of the modes in question?} It was
recently found that pressure and other non-geodesic effects have stronger
influence on the frequencies of some modes than previously thought \cite{Sra:2005:ASN:, BlaSra:2006:UNP:}. If this question is
answered, one may be able to precisely determine characteristics of
strong gravity, in particular the black hole spin (see Section \ref{subsect:measuring}).

\item {\it How are the modes excited? What is the energy source that feeds
the resonance?}  It was suggested that the excitation in the neutron star
case could be by a direct influence of the neutron star spin (e.g. due to
a magnetic coupling) \cite{Lee-etal:2004:APJ:,Klu-etal:2004:APJ:}, and in the black hole case by influence of the
turbulence \cite{Abr:2005b:AN:,Bra:2005:AN:,Vio-etal:2006:AA:}. Understanding of this point may provide strong constrains
on the magnitude and nature of turbulence in accretion disks.

\item {\it How is the X-ray flux modulated by accretion disk
oscillations?} It was suggested that in the case of neutron star sources
modulation occurs at the boundary layer \cite{Hor:2005:ASN:}, and in the black hole sources it is due to gravitational lensing and Doppler effect --- at least partially \citep[see, e.g.,][]{Bur-etal:2004:ASTRJ2L:, Bur:2005:ASN:}.
The mechanism of modulation must be related to connection between QPOs and the ``spectral states'' \cite{Rem:2005:ASN:}.

\end{itemize}

\section{Black hole high frequency QPOs}
\label{sect:BHs}

\subsection{The 3:2 ratio and $1/M$ scaling}

\begin{figure}[t]
\begin{minipage}{\hsize}
\begin{center}
\includegraphics[width=.95\textwidth]{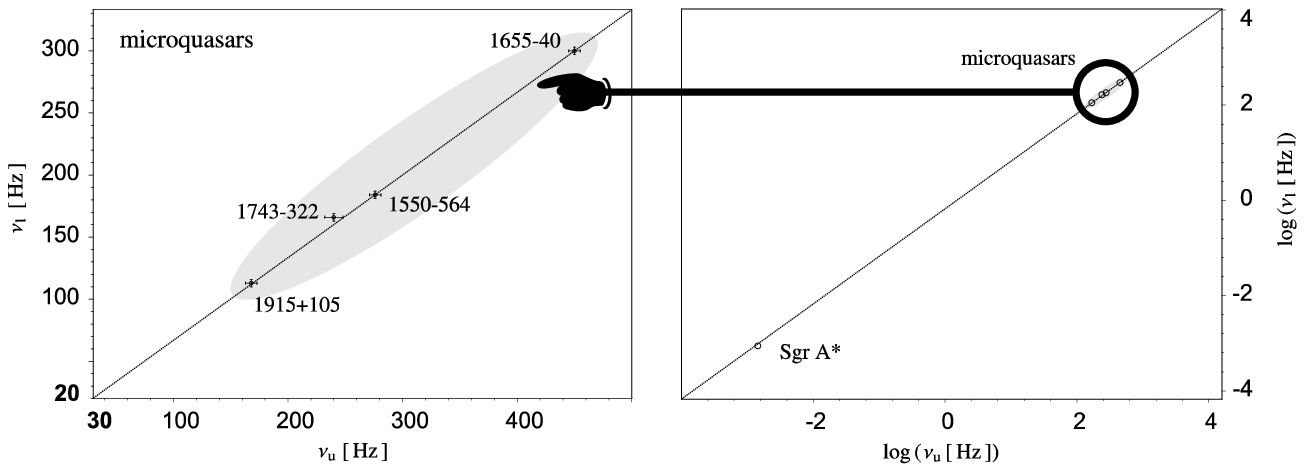}
\end{center}
\caption{The microquasar kHz QPOs frequencies in a 3:2 ratio \cite{McCRem:2004:CXS:} may be extrapolated to the black hole in the Galactic center \cite{Asch-etal:2004:AA:, Tor:2005:AA:}.}
\end{minipage}
\label{fig:3:2}
\end{figure}

\begin{figure}
\begin{minipage}{0.5\hsize}
~~~~~~
\includegraphics[width=.9\textwidth]{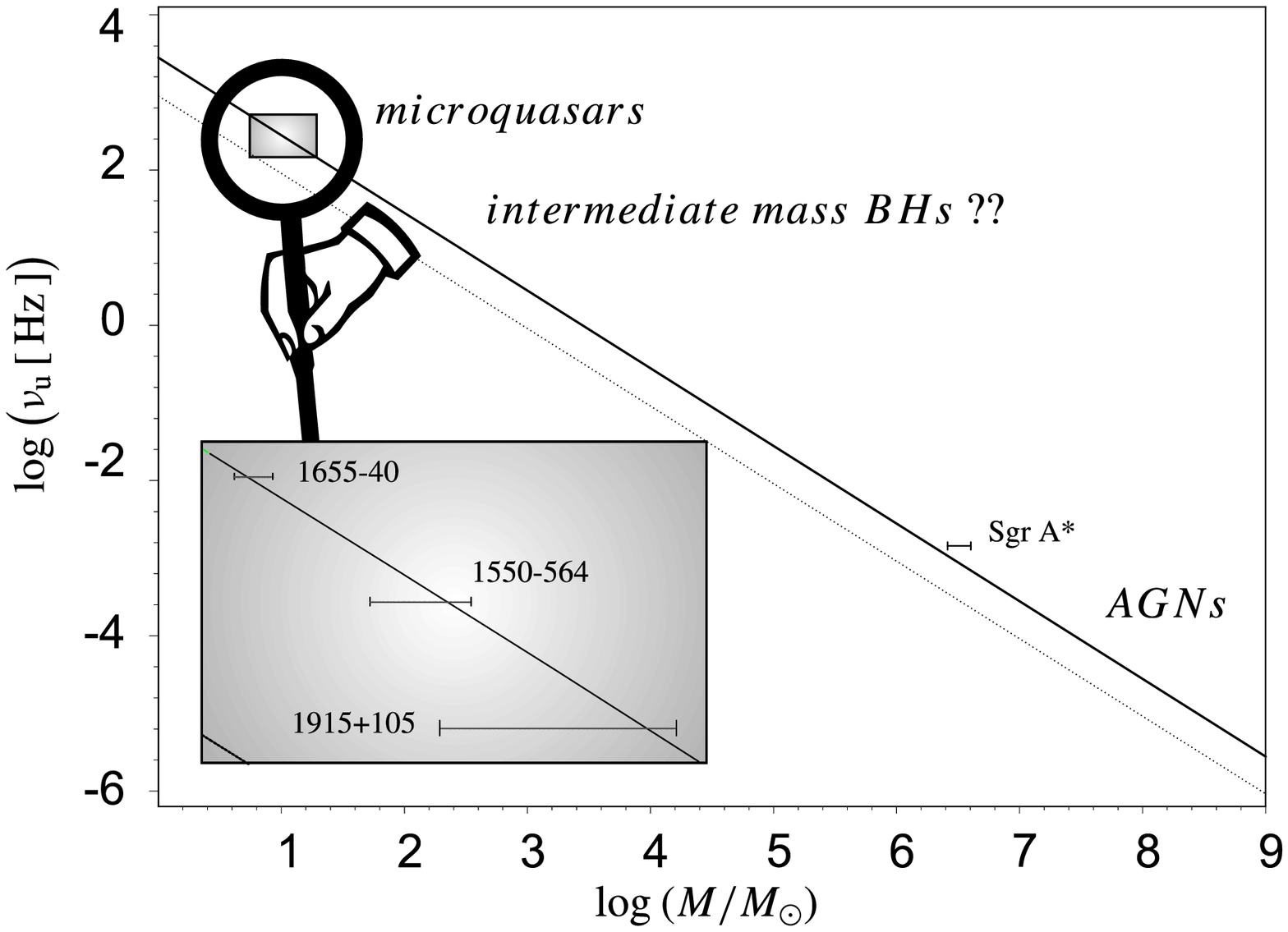}
\hfill
\end{minipage}
~~~~
\begin{minipage}{0.5\hsize}
\includegraphics[width=0.8\textwidth]{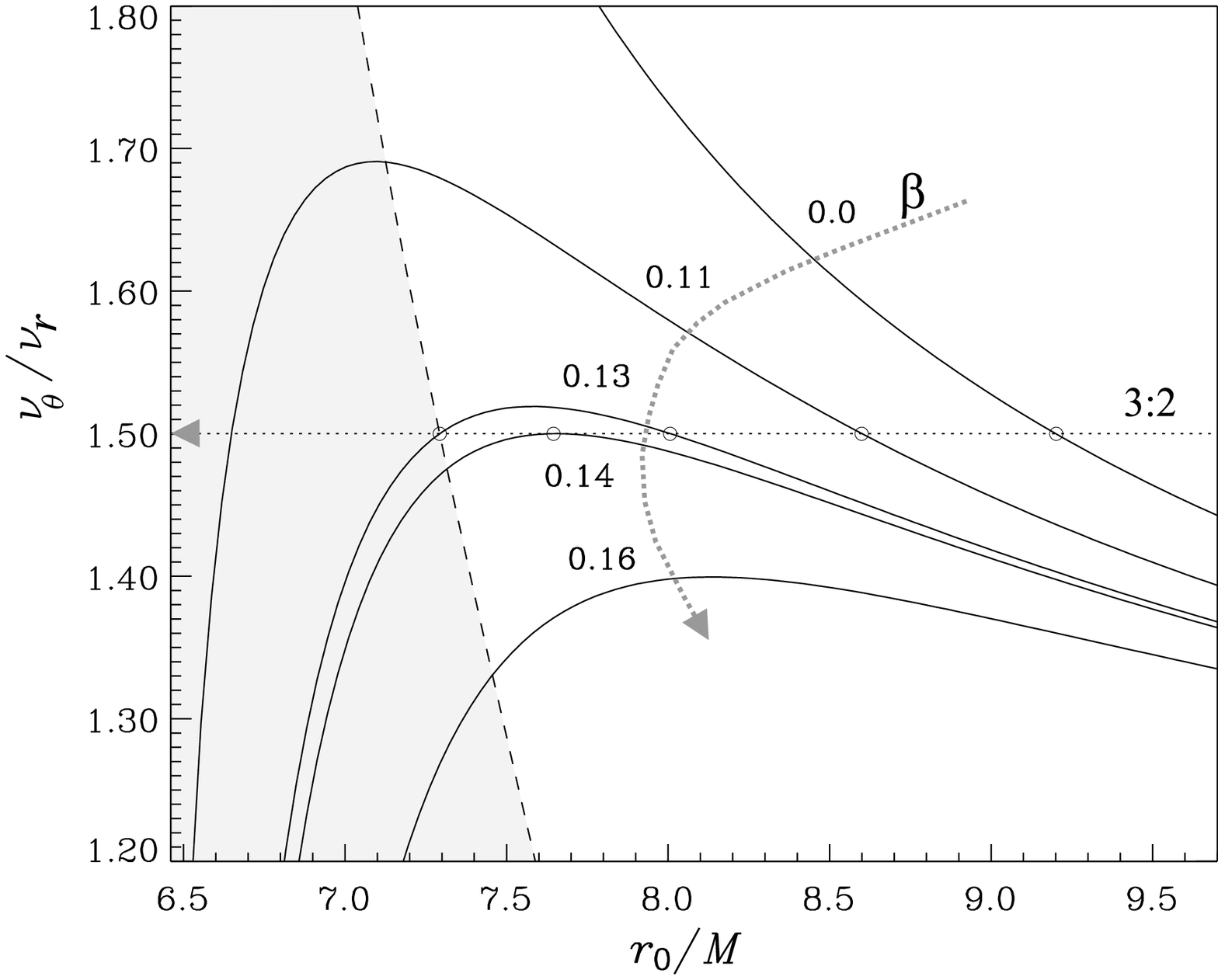}
\hfill
\end{minipage}
\caption{Left: 1/M scaling in black hole sources. The solid line is the best $1/M$ fit for microquasars by \citet{McCRem:2004:CXS:} and possibly identified with the 3:2 epicyclic resonance frequency for $a\sim0.97$, dotted line is plotted for the same resonance and $a=0$. Right: Position of the center of a torus fulfilling the 3:2 resonance condition, as a function of the the $\beta$  parameter charecterizing non-slenderity of the torus \cite{BlaSra:2006:UNP:}. In this Pseudo-Newtonian result, the resonant frequencies increase with $\beta$. A similar effect in the Kerr metric would lower the black hole spin parameter estimated for a given QPO frequency.}
\label{fig:1M}
\end{figure}

The 3:2 ratio is a strong argument in favor of explaining the QPOs in terms of a resonance \citep[see, e.g.,][for details]{AbrKlu:2004:AIP:}.
%%%%%%%%%%%%%%%%%%%%%%%%%%%%%%%%%%%%%%%%%%%%%%%%%%%%%%%%%%%%
%%%%%%%%%%%%%%%%%%table: observed \nu and M %%%%%%%%%%%%%%%%
%%%%%%%%%%%%%%%%%%%%%%%%%%%%%%%%%%%%%%%%%%%%%%%%%%%%%%%%%%%%
\begin{table}[b!]
\begin{tabular}{ l  c  c  c  c  c  l }
            \hline
\tablehead{1}{l}{b}
{Source\tablenote{Twin peak QPOs first reported by \cite{Str:2001:APJ:,Rem-etal:2002:APJ:,Hom-etal:2003:ATEL:,Rem-etal:2003:APJ:,Asch-etal:2004:AA:}.}
}
&
\tablehead{1}{c}{b}{$\nu_{\rm{u}}\,$[Hz]}&
\tablehead{1}{c}{b}{$\Delta\nu_{\mathrm{upp}}\,$[Hz]}&
\tablehead{1}{c}{b}{$\nu_{\rm {l}}\,$[Hz]}&
\tablehead{1}{c}{b}{$\Delta\nu_{\rm{l}}\,$[Hz]}&
\tablehead{1}{c}{b}{$2\nu_{\rm{u}}/3\nu_{\rm{l}}- 1$}&
\tablehead{1}{c}{b}
{Mass\tablenote{See \cite{Gre-etal:2001:APJ:,Oro-etal:2002:APJ:,Gre-etal:2001:NAT:,McCRem:2004:CXS:} for the microquasars. Note that there is also the different estimate for GRO~1655--40: $M=(5.4 \pm 0.3)\, {\rm M}_\odot$~\cite{BeePod:2002:MNRAS:}. 
}
~~[\,M$_{\odot}$\,]}\\
\hline
GRO~1655--40    & 450&$\pm\,3$& 300&$\pm\,5$ &\phantom{-}0.000& $\phantom{1}$6.0\, --- $\phantom{1}$6.6   \\
XTE~1550--564   & 276&$\pm\,3$& 184&$\pm\,5$& \phantom{-}0.000& $\phantom{1}$8.4\, --- 10.8   \\          
\phantom{RS}H~1743--322   & 240&$\pm\,3$& 166&$\pm\,8$&-0.036& not measured    \\
GRS~1915+105    & 168&$\pm\,3$& 113&$\pm\,5$& \phantom{-}0.009& 10.0 \,--- 18.0   \\
\hline
\end{tabular}
\caption{Frequencies of twin peak QPOs in microquasars.}
\label{table:1}
\end{table}
In general relativistic phenomena around a mass $M$, all frequencies scale as $1/M$. This agrees with black holes having lower QPO frequencies than neutron  stars and is suggestive of a relativistic origin for kHz QPOs [1].
 \citet{McCRem:2004:CXS:} found that in microquasars the frequencies do indeed scale inversely with mass:
\begin{equation}
\label{eq:bestfit}
\nu_{u}= 2.793 \left ( {M_0 \over M_{\odot}}\right )^{-1}\, \mathrm{kHz}.
\end{equation}
The $1/M$ scaling could be used to estimate the black hole mass in AGNs and ULXs \cite{Abr-etal:2004:APJ:}, if HF QPOs were to be discovered in those sources (see Figures \ref{fig:3:2}, \ref{fig:1M}).     
%%%%%%%%%%%%%%%%%%%%%%%%%%%%%%%%%%%%%%%%%%%
%%%%%%%%%%%%%%%%%%%%%%%%%
%%%%%% 3:2 figure %%%%%%%
%%%%%%%%%%%%%%%%%%%%%%%%%
%%%%%%%%%%%%%%%%%%%%%%%%%
%%%%%%%%%%%%%%%%%%%%%%%%%
%%%%%%%%%%%%%%%%%%%%%%%%%
%%%%%% 1/M and Evas fig.
%%%%%%%%%%%%%%%%%%%%%%%%%
%%%%%%%%%%%%%%%%%%%%%%%%%
%%%%%%%%%%%%%%%%%%%%%%%%%

%%%%%%%%%%%%%%%%%%%%%%%%%%%%%%%%%%%%%%%%%%%%%%%%%%%%%%%%%%%%%%
%%%%%%%%%%% MEASURING THE SPIN %%%%%%%%%%%%%%%%%%%%%%%%%%%%%%%
%%%%%%%%%%%%%%%%%%%%%%%%%%%%%%%%%%%%%%%%%%%%%%%%%%%%%%%%%%%%%%
\subsection{Measuring the black hole spin}
\label{subsect:measuring}

According to the \emph{epicyclic} resonance hypothesis, the two modes in resonance have eigenfrequencies $\nu_{\rm rad}$ (equal to the radial epicyclic frequency) and $\nu_{\rm v}$ (equal to the vertical orbital frequency $\nu_{\theta}$ or to the Keplerian frequency $\nu_{\rm K}$). Several resonances of this kind are possible, and have been  discussed \citep[see, e.g.,][]{AbrKlu:2004:AIP:}).

Formulae for the Keplerian and epicyclic frequencies $\nu_{\rm vert}$ and $\nu_{\rm rad}$ in the gravitational field of a rotating Kerr black hole with the mass $M$ and internal angular momentum $a$ (\emph{spin} $a$) are well known \cite{NowLeh:1998:TAD:}:

\begin{eqnarray}
\label{eq:epfrequencies}
\nu_{\rm r}^2 &=& {\nu_{\rm K}^2}\,\left(1-6\,x^{-1}+ 8 \,a \, x^{-3/2} -3 \, a^2 \, x^{-2} \right),
~~~~~~~~~~~~~~~~~~~~~~~~~~~~~~~\,
x = \frac{r}{r_{G}}=\frac{r}{2GM/c^2}, 
\nonumber\\
\nu_{\theta}^2 &=& {\nu_{\rm K}^2}
\,\left(1-4\,a\,x^{-3/2}+3a^2\,x^{-2}\right),
~~~~~~~~~~~~~~~~~~~~~~~~~~~~~~~~~~~~~~~~~~~~
\nu_{\mathrm{K}} = {1\over 2\pi}\left ({{GM_0}\over {r_G^{~3}}}\right )^{1/2}\left( x^{3/2} + a \right)^{-1}\,.
\end{eqnarray}

For a particular resonance n:m, the equation
\begin{equation}
\mathrm{n}\nu_{\rm r} = \mathrm{m}\nu_{\rm v}; \quad \nu_{\rm v}=\nu_\theta\,~\mathrm{or}~\,\nu_\mathrm{K}
\end{equation}
determines the dimensionless resonance radius $x_{\mathrm{n}:\mathrm{m}}$ as a function of spin $a$.
Thus, from the observed frequencies and from the estimated mass one can calculate the relevant spin of a central black hole \citep{AbrKlu:2001:AA:,Tor-etal:2005:AA:}.

Results of this procedure were summarized, e.g., in \cite{Tor:2005:ASN:}. Several resonance models give the values of spin in the range $a\in (0,1)$. In particular, the 3:2 epicyclic parametric (or internal) resonance model supposed to be most natural in the Einstein gravity \cite{Hor:2005:ASN:} implies microquasar black hole spin $a\sim 0.9$. However, the most recent results of the spin estimate for GRO~1655--40 \citep[e.g.,][]{Sha-etal:2006:APJ:} claims
the spin to be $\sim 0.7$. It could be interesting that recently proposed 3:2 periastron precession resonance \cite{Bur:2005:RAG:} implies  the spin of GRS 1915+105 to be also $a\sim 0.7$.

On the other hand, \emph{above mentioned resonance spin estimates are based on a resonance between epicyclic frequencies equal to those for exactly geodesic motion} (\ref{eq:epfrequencies}). Recently \cite{Sra:2005:ASN:, BlaSra:2006:UNP:}, it was found that pressure and other non-geodesic effects have stronger
influence on the frequencies than previously thought. At the moment, it is difficult to give plausible estimate how this effects change the spin predictions as the theory is not developed enough, but work on this is in progress. A rough estimate, based on Pseudo-Newtonian calculations of \cite{BlaSra:2006:UNP:} and the assumption of the Kerr case being qualitatively same, says that corrected values of spin required from 3:2 parametric resonance model will be rather lower then was supposed (see Fig. \ref{fig:1M}, right panel).

\section{Neutron star high frequency QPOs}
\label{sect:NS}

As was shortly reminded above, for the black holes  $\nu_\mathrm{u}$ and $\nu_\mathrm{l}$ appear to be fairly fixed having the well defined ratio $\nu_\mathrm{u}/\nu_\mathrm{l} = 3/2$. Contrary to this, $\nu_\mathrm{u}$, $\nu_\mathrm{l}$ in neutron star sources vary by hundreds of Hertz, along ``Bursa lines''
%----------------------------------------
\begin{equation} 
  \label{eqn:bursa} 
  \nu_\mathrm{u} = A\,\nu_\mathrm{l} + B.
\label{eq:bursaline}
\end{equation}
%----------------------------------------
%%%%%%%%%%%%%%%%%%%%%%%%%
%%%%%% Bursaplot
%%%%%%%%%%%%%%%%%%%%%%%%%
\begin{figure}[h!]
\begin{minipage}{1\hsize}
\begin{center}
\includegraphics[width=.49\textwidth]{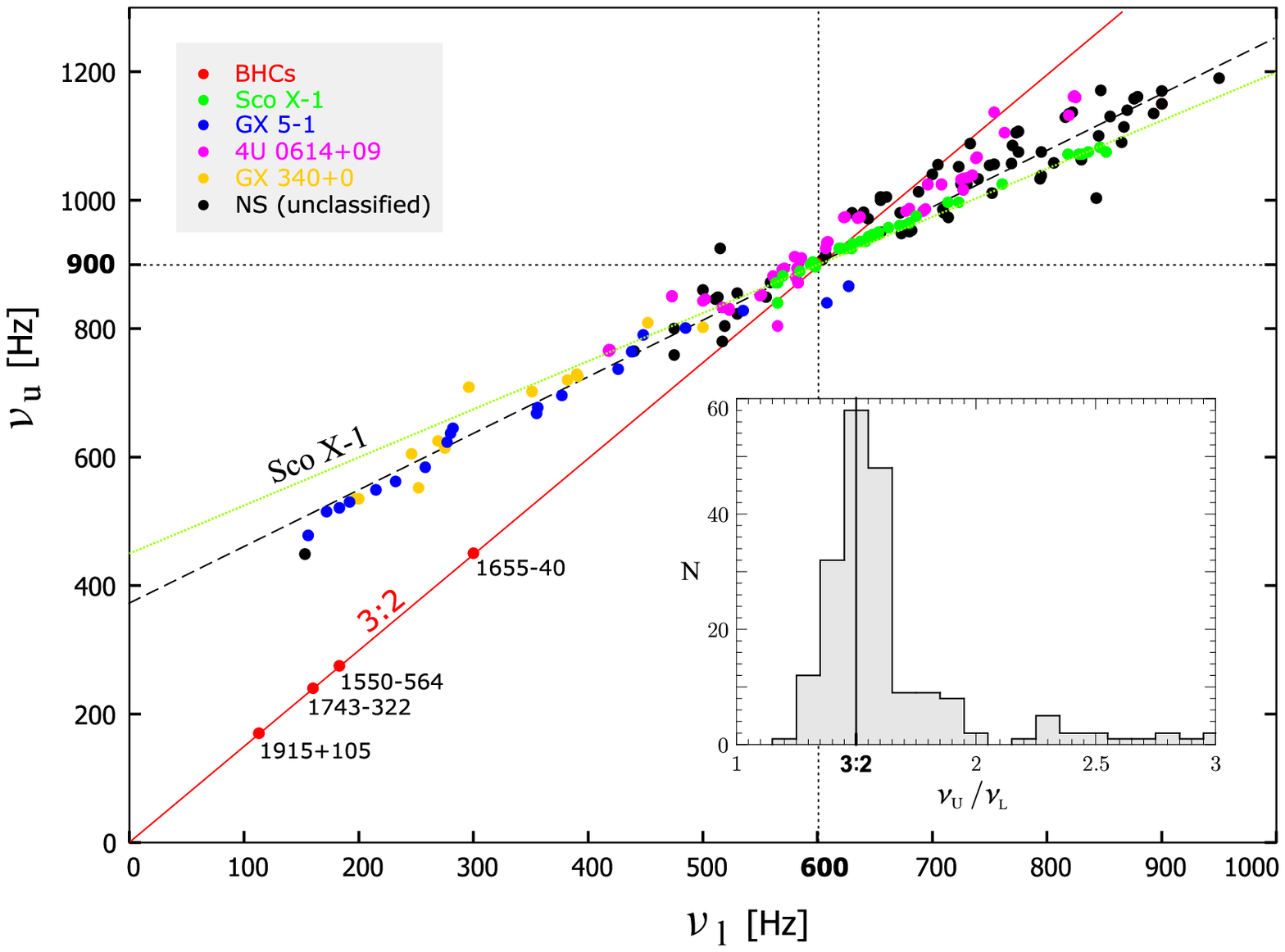}
\end{center}
\end{minipage}
\caption{Bursa plot: The linear fit of the neutron-stars data (dashed line) is obviously inconsistent with the 3:2 relation, but the frequency ratio is peaked to the 3:2 value, i.e. close to the 3:2 resonance \cite{Abr-etal:2003:AA:}. \citet{Psa-etal:1998:ASTRJ2L:} has shown, in a different context, that frequency pairs of all sources cluster along a single line. Bursa (2002, unpublished) also pointed that each source follows a slightly different line. It was demonstrated by \cite{Abr-etal:2003:PUBASJ:, Reb:2004:PUBASJ:, Hor:2004:RAGtime4and5:Proceedings:} that variations of $\nu_\mathrm{u}$ and $\nu_\mathrm{l}$ along the line (\ref{eqn:bursa}) can be explained within the non-linear resonance model for QPOs (see Section \ref{subsect:anti}).}
\label{fig:Bursa}
\end{figure}
%%%%%%%%%%%%%%%%%%%%%%%%%
%%%%%%%%%%%%%%%%%%%%%%%%%

\subsection{Ratio vs. frequency distribution}

The distribution of the $\nu_\mathrm{u}/\nu_\mathrm{l}$ frequency ratio in neutron star sources was shown by \citet{Abr-etal:2003:AA:} to cluster near the 3:2 value. This result was later criticised by \citet{Bel-Men-Hom:2005:ASTRA:} who thought that the relevant quantity must be the distribution of a single frequency.

\begin{figure}[h!]
\begin{minipage}{.85\hsize}
\includegraphics[width=1\textwidth]{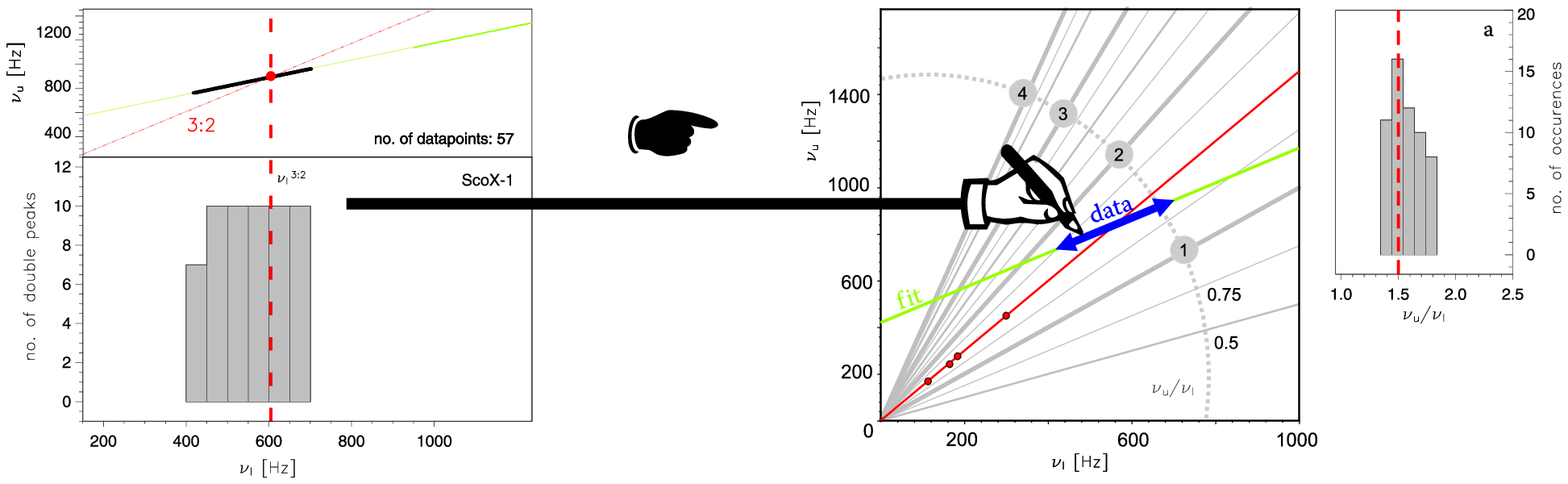}
\end{minipage}
\caption{A simulated segment of 57 frequency-frequency points (related by $\nu_\mathrm{u}=0.75\nu_\mathrm{l}+450$) with a flat distribution in the lower frequency is placed approximately at the position of Sco X-1 data in $\nu$-$\nu$ diagram. Resulting histogram of ratio is peaked at the 3:2 value.}
\label{fig:scox1a}
\end{figure}

\begin{figure}[t!]
\begin{minipage}{1\hsize}
\includegraphics[width=1\textwidth]{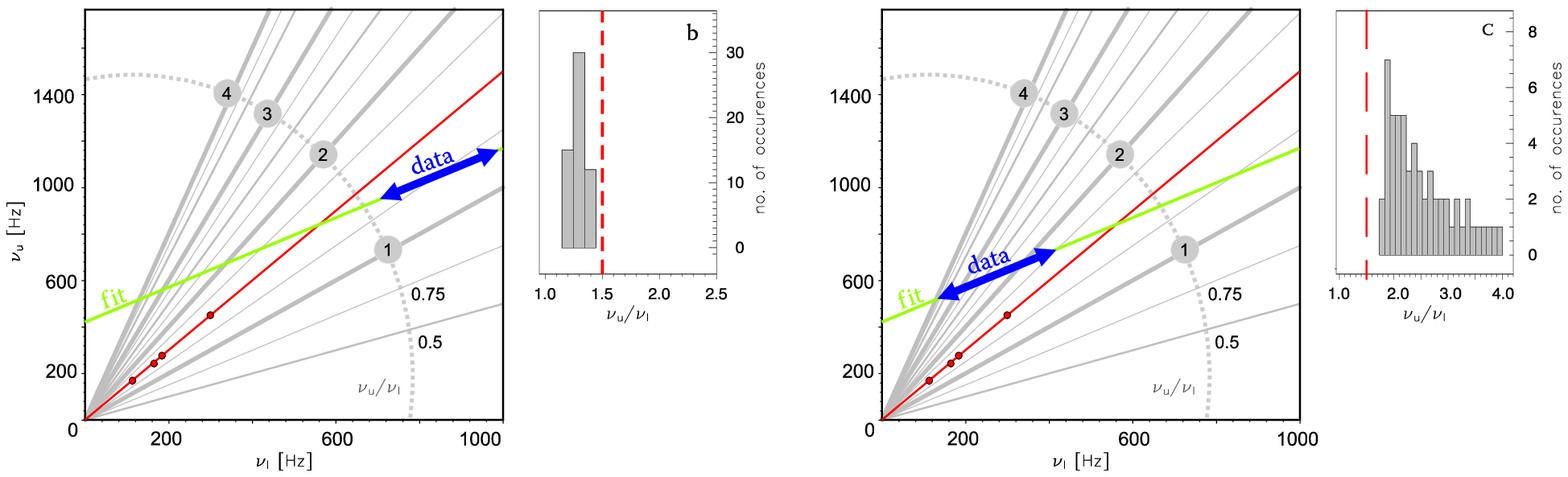}
\end{minipage}
\caption{Displacing the data segment from its position consistently with the same fixed relation $\nu_\mathrm{u}=0.75\nu_\mathrm{l}+450$ results in completely different histogram even if the displacement is of the order of length of datasegment.}
\label{fig:scox1b}
\end{figure}

We stress that this is a misunderstanding and these two distributions touch different problems. Linear \emph{correlation between the frequencies doesn`t imply the ratio clustering} because for this the location of data segment is important as well (see Figures \ref{fig:scox1a} and \ref{fig:scox1b}). Nevertheless, \citet{Bel-Men-Hom:2005:ASTRA:} confirmed the 3:2 clustering in the neutron star sources and shown also the very interesting clustering close to the related resonant frequency.
\bigskip

%%%%%%%%%%%%%%%%%%%%%%%%%
%%%%%% scox1a
%%%%%%%%%%%%%%%%%%%%%%%%%

%%%%%%%%%%%%%%%%%%%%%%%%
%%%%%% scox1b
%%%%%%%%%%%%%%%%%%%%%%%%%
%%%%%%%%%%%%%%%%%%%%%%%%%
%%%%%%%%%%%%%%%%%%%%%%%%%

\subsection{Anticorrelation between slope and shift}
\label{subsect:anti}

\citet{Abr-etal:2005:RAGtime6and7:} have shown that linear relations (\ref{eq:bursaline}) for twelve neutron star sources are anticorrelated. Below we shortly remind the connection between this anticorrelation and properties of weakly coupled nonlinear oscillators \citep[see][for details]{Abr-etal:2006:MNRAS:}.

The frequencies of non-linear oscillations may be written in the form
%----------------------------------------
\begin{eqnarray}
  \nu_\mathrm{l} &=& \nu_{\mathrm{l}}^0 + \Delta\nu_\mathrm{l}\,, \quad
  \Delta\nu_\mathrm{l}={\nu}_\mathrm{l}^{0}(\kappa_\mathrm{l} a_\mathrm{l}^2 + \kappa_\mathrm{u} a_\mathrm{u}^2)\,,
  \label{eq:corrections}
\nonumber\\
  \nu_\mathrm{u} &=& \nu_{\mathrm{u}}^0 + \Delta\nu_\mathrm{u}\,,
  \quad
  \Delta\nu_\mathrm{u}={\nu}_\mathrm{u}^{0}(\lambda_\mathrm{l} a_\mathrm{l}^2 + \lambda_\mathrm{u} a_\mathrm{u}^2)\,,
\end{eqnarray}
%----------------------------------------
where $\nu_\mathrm{l}^0$ and $\nu_\mathrm{u}^0$ are the eigenfrequencies of oscillator and $a_l$, $a_u$ are the amplitudes of oscillations.

If the two amplitudes, $a_l$ and $a_u$ are correlated, i.e.,
if they are functions of a single parameter $s$, one gets a linear relationship
between $\nu_\mathrm{u}$ and $\nu_\mathrm{l}$, obtained from lowest order expansion
\footnote{Note that if one includes more terms in the expansion, the frequency-frequency correlation could deviate from straight line.}
 in the parameter $s$.
 
The inferred coefficients $A$ and $B$ of linear relation (\ref{eq:bursaline}) will vary from source to source, but if  the eigenfrequencies of oscilations ($\nu_\mathrm{l}^0$, $\nu_\mathrm{u}^0$) are universal for the sources (being fixed by the common space-time metric and mass), then these linear relations themselves will satisfy
%----------------------------------------
\begin{equation}
  A = \frac{\nu_{\mathrm{u}}^0}{\nu_{\mathrm{l}}^0}-\frac{1}{\nu_{\mathrm{l}}^0}\,B\, = A_0 - \frac{1}{\nu_{\mathrm{l}}^0}\,B\,.
  \label{eq:anticorr3}
\end{equation}
%----------------------------------------
Therefore, \emph{the slope $A$ and the shift $B$ are anticorrelated}.
If the eigenfrequencies of oscillators slightly differ of factor $\sigma$ (e.g., due to the difference of the mass and connected $1/M$ frequency scaling) but the ratio of eigenfrequencies is fixed, this anticorrelation reads
%----------------------------------------
\begin{equation}
  A = A_0 - \frac{1}{\sigma\nu_{\mathrm{l}}^0}\,B\,.
  \label{eq:anticorr4}
\end{equation}
%----------------------------------------

\begin{figure}[h]
\begin{minipage}{1\hsize}
\centering
\includegraphics[width=0.48\textwidth]{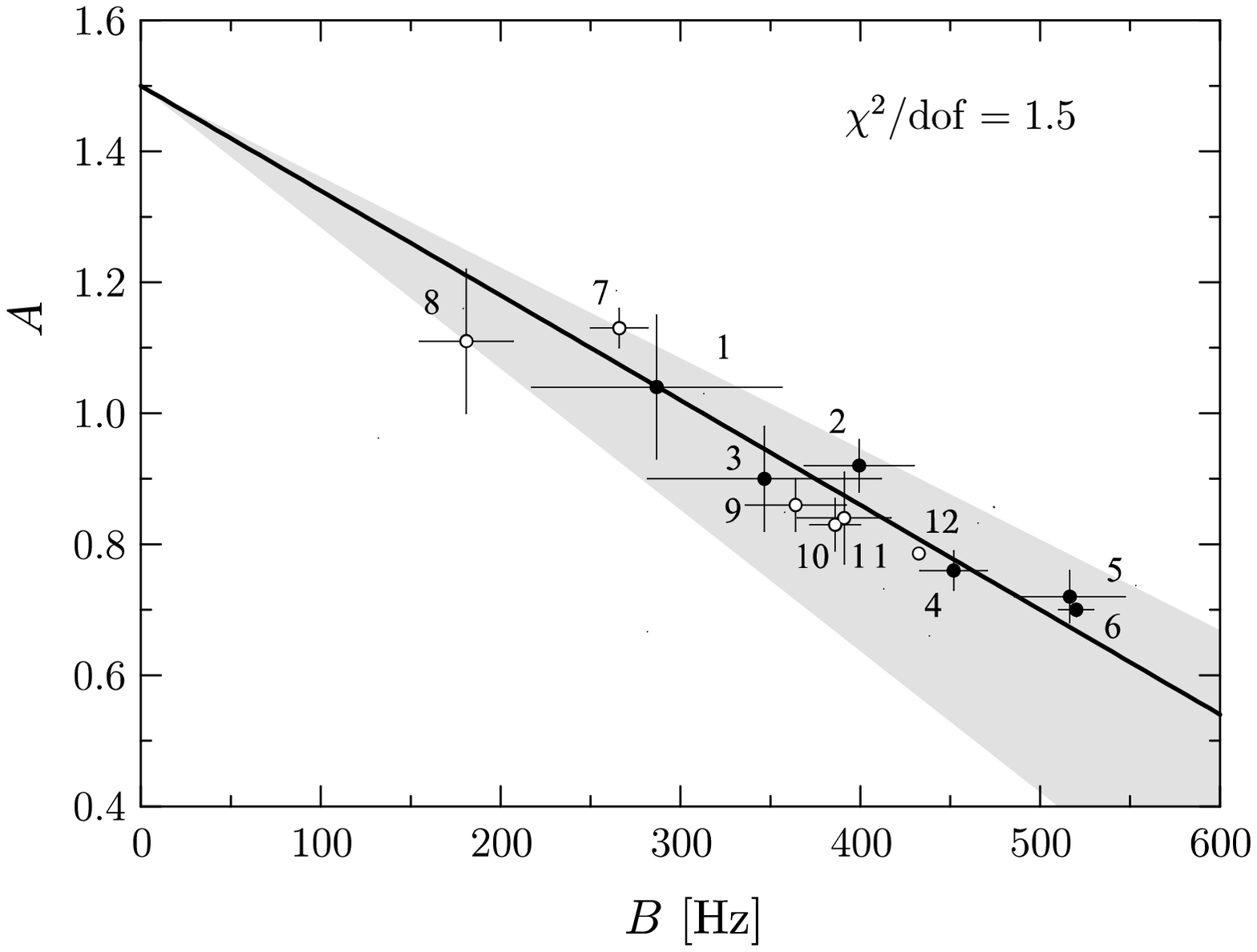}
\end{minipage}
\caption{The coefficients of the linear frequency-frequency relation for kHz QPOs in twelve neutron-star sources:
\newline {1)~4U~0614, 2)~4U~1728, 3)~4U~1820, 4)~4U~1608, 5)~4U~1636, 6)~4U~1735 \cite{Abr-etal:2006:MNRAS:}; 7)~4U~1915 \cite{Boi-etal:2000:}; 8)~XTE~J1807~\cite{Lin-etal:2005:APJ:}; 9)~GX~17+2~\cite{Hom-etal:2002:ASTRJ2:}; 10)~GX 34+0 \cite{Jon-etal:2000:ASTRJ2:}; 11)~GX 5-1 \cite{Jon-etal:2002:MONNR:}; 12)~Sco X-1 \cite{Bel-Men-Hom:2005:ASTRA:}.}\newline
The data are consistent with a linear relationship intercepting the $A$ axis at the value 1.5 (Note good quality of fit by straight line). The scatter (shaded area) expressed in terms of $\sigma$ coefficient from eq. (\ref{eq:anticorr4}) is most likely  from  the interval $\sigma\in(0.8, 1.2)$. Under the assumption that difference in the eigenfrequencies is caused by different neutron star mass, the mass should differ of factor $\sim$1.5.
}
\label{fig:anti}
\end{figure}

Thus, for given type of resonance, the individual pairs $A, B$ should locate on lines inside a triangle  in the $A$-$B$ plane with a quite well determined vertex at $[0,\,A_0]$, and with the size of its base proportional to the scatter in ${\nu}_\mathrm{l}$. In particular for the 3:2 resonance, the vertex should be $[0,1.5]$. Figure \ref{fig:anti} suggests that this is indeed the case.
\bigskip

In principle, for the \emph{3:2 parametric epicyclic resonance model} the eigenfrequencies of oscillations can be directly identified with the epicyclic frequencies of Keplerian motion at the location of this 3:2 resonance. This implies that the \emph{generic} neutron star mass connected to the straight line in Fig. \ref{fig:anti} is about $1M_\odot$ (first noticed by Bursa 2003, unpublished). However, recent results \cite{BlaSra:2006:UNP:} suggest that due to a pressure and other non-geodesic effects, this mass will be higher.

\newpage

%%%%%%%%%%%%%%%%%%%%%%%%
%%%%%% anti
%%%%%%%%%%%%%%%%%%%%%%%%%
%%%%%%%%%%%%%%%%%%%%%%%%%
%%%%%%%%%%%%%%%%%%%%%%%%%

\subsection{RMS amlitude evolution across the resonance point}

\begin{figure}[h!]
\begin{minipage}{0.5\hsize}
\centering
\includegraphics[width=0.8\textwidth]{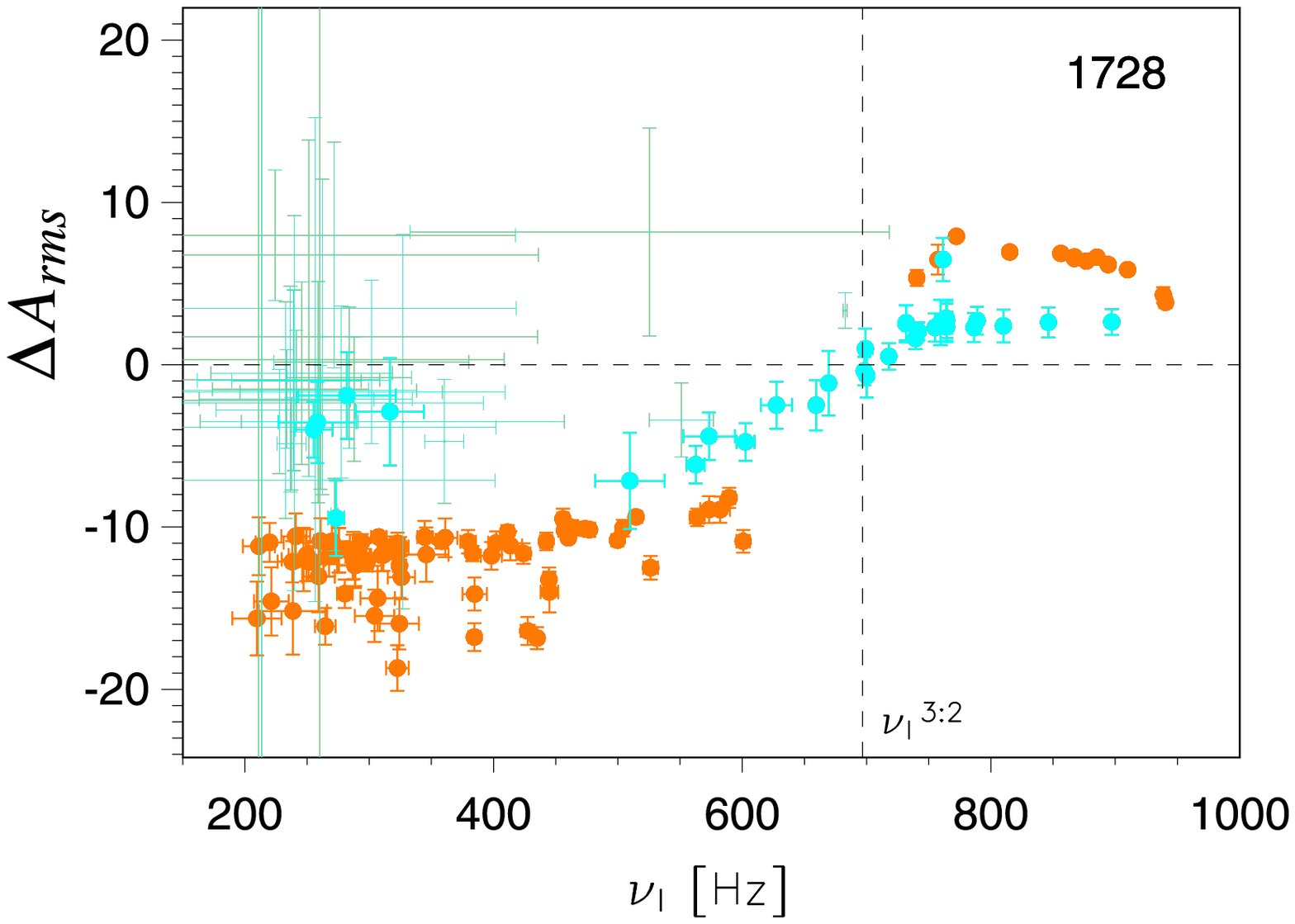}
\end{minipage}
\begin{minipage}{0.5\hsize}
\centering
\includegraphics[width=0.8\textwidth]{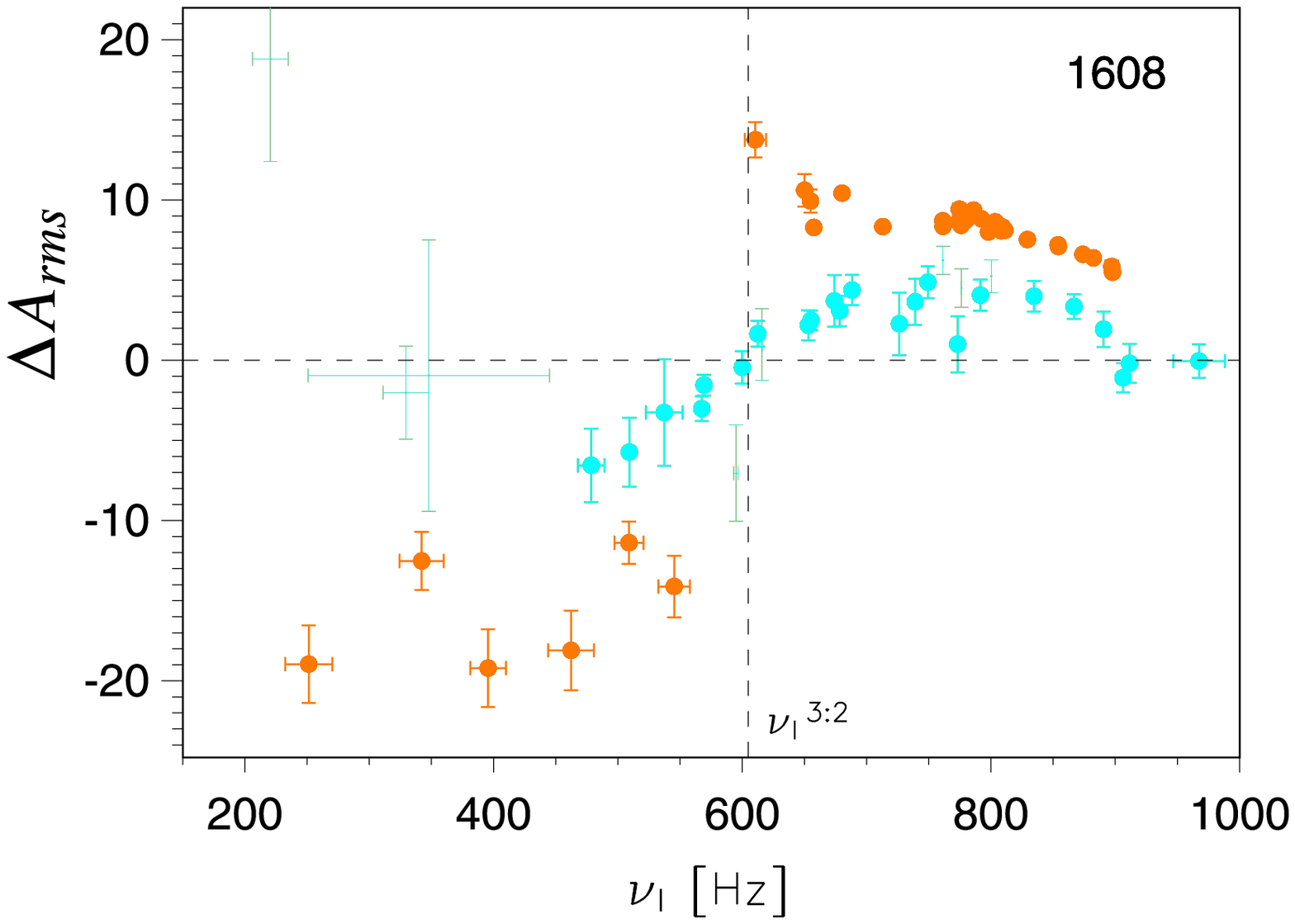}
\end{minipage}
\caption{The difference of the rms amplitudes vs. lower QPO frequency for two neutron star atoll sources 4U 1728-34 and 4U 1608-52. Each datapoint corresponds to one continuous segment of observation (error bars without central circle denote insignificant observations). Cyan points correspond to the difference in double peaks. Orange points follow from single QPOs identified through the Q-factor \citep[see][]{Bar-etal:2005:AN:}, their counterparts are here supposed (probably not quite realistically) to have rms equal zero. It seems that up to the 'resonance point' the amplitudes are larger in the case of the upper frequency, equal when 'Bursa line' pass 3:2 ratio and above this, the lower QPO has the stronger amplitude. \emph{Based on \cite{Tor-etal:2006:INPR:} in preparation.}}
\label{fig:16081728}
\end{figure}

\begin{figure}[t]
\begin{minipage}{0.5\hsize}
\centering
\includegraphics[width=0.84\textwidth]{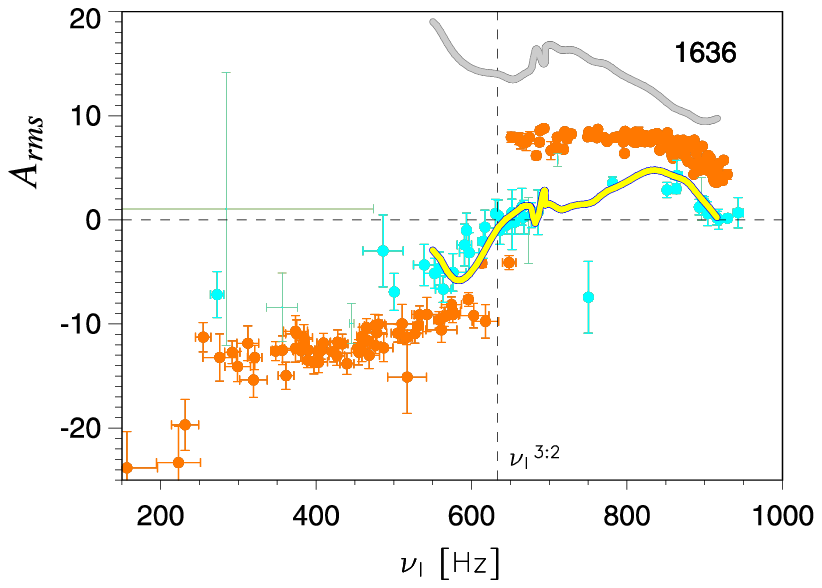}
\end{minipage}
\begin{minipage}{0.5\hsize}
\centering
\includegraphics[width=0.84\textwidth]{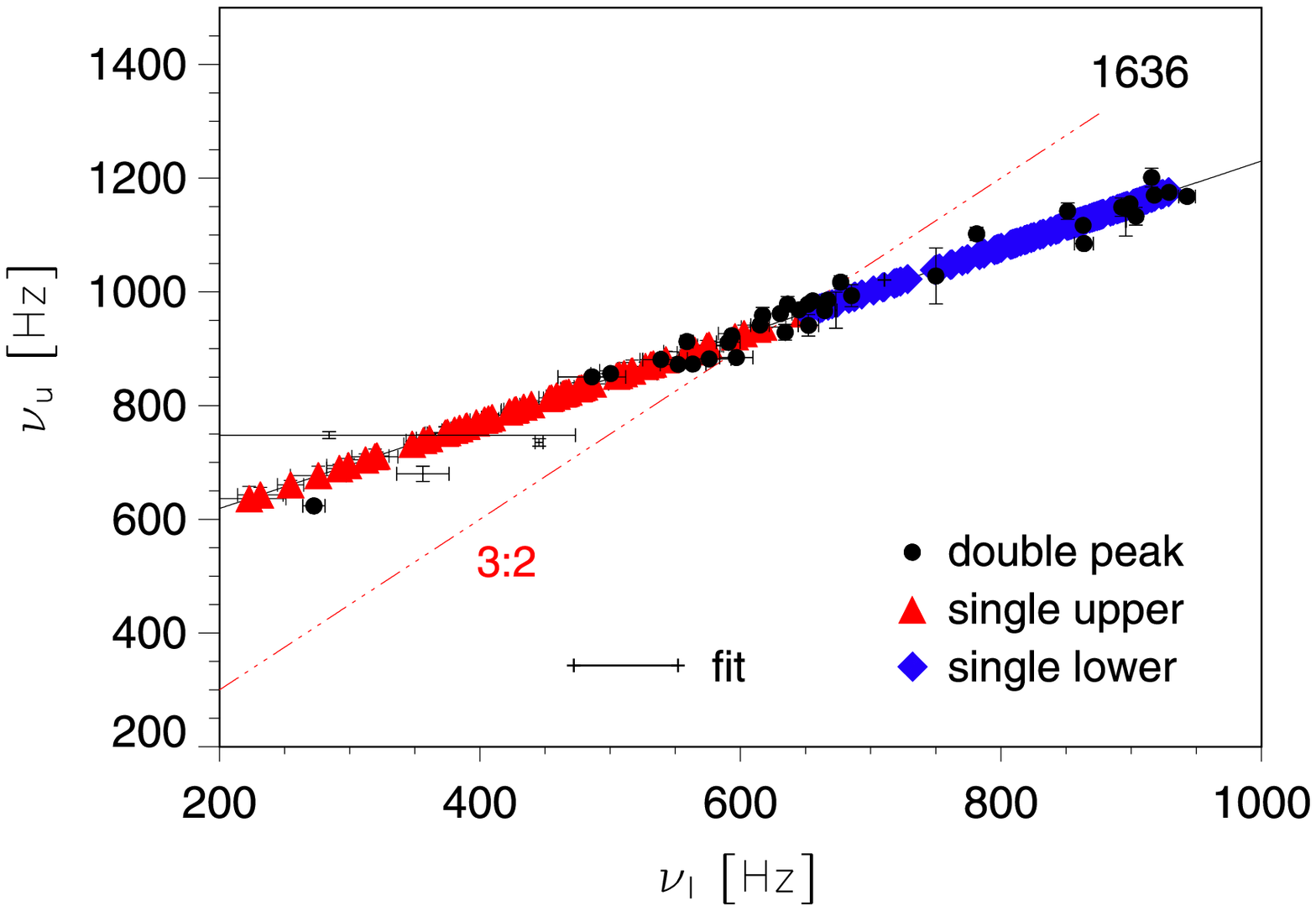}
\end{minipage}
\caption{Left: The difference (yellow curve and symbols as in previous figure) and total (grey curve) of the rms amplitudes for atoll source 4U 1636-53. The continuous curves are calculated from polynomial interpolation of the values of rms \cite{Bar-etal:2005:MNRAS:} for the upper and lower frequency obtained through shift and add over all PDS. Right: Double peaks vs. single QPO which are distinguished using Q-factor and placed in accord to linear fit. Up to the intersection of the fit with the 3:2 line, lower frequencies are often supressed (i.e., difficult to find) as the single 'upper' frequency  dominate, while above this the situation is opposite. \emph{Based on \cite{Tor-etal:2006:INPR:} in preparation.}}
\label{fig:1636}
\end{figure}

Quite recently, T\"or\"ok discovered the strong change in character of the observed QPO amplitudes (\citet{Tor-etal:2006:INPR:}). The difference between the rms amplitude of the lower QPO and the rms amplitude of the relevant upper QPO, i.e., the quantity
\begin{equation}
\Delta\,A_{rms}(\nu) \equiv A_{rms}^{\nu_\mathrm{l}}(\nu) - A_{rms}^{\nu_\mathrm{u}}(\nu),
\quad\mathrm{where}~\nu = \nu_\mathrm{l}
\end{equation}
is rather well correlated with frequency and changes its sign across the 3:2 resonance line which is illustrated in Figures~\ref{fig:16081728} and \ref{fig:1636}.

It was noticed by Hor\'ak that such phenomenon resembles general behavior of mechanical systems crossing an
internal resonance (\citet{Hor-etal:2006:INPR:}). As some frequencies approach a rational ratio the
modes involved in the resonance start to exchange energy. The slow
motion through the resonance perturbs the strict balance between
both directions of energy exchange and finally causes increase of one amplitude at the expense of the other one.
\bigskip

%%%%%%%%%%%%%%%%%%%%%%%%%
%%%%%% 1608 and 1728, 1636
%%%%%%%%%%%%%%%%%%%%%%%%%
%%%%%%%%%%%%%%%%%%%%%%%%%%%%%%%%%%%%%%%%%%%%

%%%%%%%%%%%%%%%%%%%%%%%%%%%%%%%%%%%%%%%%%%%%%%%%
%% BACKMATTER
%%%%%%%%%%%%%%%%%%%%%%%%%%%%%%%%%%%%%%%%%%%%%%%%

\begin{theacknowledgments}
  M.A.A., Z.S., and G.T. were supported by the Czech grant MSM 4781305903 and WK by KBN grant 2P03D01424. We thank Didier Barret, Eva \v{S}r\'{a}mkov\'{a} and Jirka Hor\'{a}k for help and discussions.  
\end{theacknowledgments}

\hyphenation{Post-Script Sprin-ger}


\begin{thebibliography}{50}
\expandafter\ifx\csname natexlab\endcsname\relax\def\natexlab#1{#1}\fi
\providecommand{\enquote}[1]{``#1''}
\expandafter\ifx\csname url\endcsname\relax
  \def\url#1{\texttt{#1}}\fi
\expandafter\ifx\csname urlprefix\endcsname\relax\def\urlprefix{URL }\fi
\providecommand{\eprint}[2][]{\url{#2}}

\bibitem[Klu{\'z}niak and Abramowicz(2000)]{KluAbr:2000:UNP:}
W.~Klu{\'z}niak, and M.~A. Abramowicz  (2000), submitted to Phys. Rev. Lett.,
  \eprint{astro-ph/0105057}.

\bibitem[van~der Klis(2005)]{Kli:2005:ASN:}
M.~van~der Klis, \emph{Astronomische Nachrichten} \textbf{326}, 798--803
  (2005).

\bibitem[Abramowicz(2005{\natexlab{a}})]{Abr:2005:ASN:}
M.~A. Abramowicz, \emph{Astronomische Nachrichten} \textbf{326}, 782--786
  (2005{\natexlab{a}}).

\bibitem[Linares et~al.(2005)]{Lin-etal:2005:APJ:}
M.~Linares, M.~van~der Klis, D.~Altamiro, and C.~B. Markwardt,
  \emph{Astrophysical Journal} \textbf{634}, 1250--1260 (2005).

\bibitem[Abramowicz and Klu{\'z}niak(2004)]{AbrKlu:2004:AIP:}
M.~A. Abramowicz, and W.~Klu{\'z}niak, \enquote{Interpreting black hole QPOs,}
  in \emph{X-ray Timing 2003: Rossi and Beyond.}, edited by P.~Karet, F.~K.
  Lamb, and J.~H. Swank, NY: American Institu of Physics, Melville, 2004, vol.
  714, pp. 21--28.

\bibitem[{\v{S}}{r\'{a}}mkov{\'{a}}(2005)]{Sra:2005:ASN:}
E.~{\v{S}}{r\'{a}}mkov{\'{a}}, \emph{Astronomische Nachrichten} \textbf{326},
  835--837 (2005).

\bibitem[Blaes et~al.(2006)]{BlaSra:2006:UNP:}
O.~Blaes, E.~\v{S}ramkov{\'{a}}, et~al.  (2006), in preparation.

\bibitem[Lee et~al.(2004)]{Lee-etal:2004:APJ:}
W.~H. Lee, M.~A.~A. M.A., and W.~Kluzniak, \emph{Ap.J.Lett.} \textbf{603}, L93
  (2004).

\bibitem[Kluzniak et~al.(2004)]{Klu-etal:2004:APJ:}
W.~Kluzniak, M.~A. Abramowicz, S.~Kato, W.~H. Lee, and N.~Stergioulas,
  \emph{ApJ L} \textbf{603}, L89 (2004).

\bibitem[Abramowicz(2005{\natexlab{b}})]{Abr:2005b:AN:}
M.~A. Abramowicz, \emph{Astronomische Nachrichten} \textbf{326}, 728
  (2005{\natexlab{b}}).

\bibitem[Brandenburg(2005)]{Bra:2005:AN:}
A.~Brandenburg, \emph{Astronomische Nachrichten} \textbf{326}, 787 (2005).

\bibitem[Vio et~al.(2006)]{Vio-etal:2006:AA:}
R.~Vio, P.~Rebusco, P.~Andreani, H.~Madsen, and R.~Overgaard, \emph{A\&A}
  (2006), in press.

\bibitem[Hor{\'{a}}k(2005)]{Hor:2005:ASN:}
J.~Hor{\'{a}}k, \emph{Astronomische Nachrichten} \textbf{326}, 845--848 (2005).

\bibitem[Bursa et~al.(2004)]{Bur-etal:2004:ASTRJ2L:}
M.~Bursa, M.~A. Abramowicz, V.~Karas, and W.~Klu{\'z}niak, \emph{ApJ L}
  \textbf{617}, L45--L48 (2004), \eprint{astro-ph/0406586}.

\bibitem[Bursa(2005{\natexlab{a}})]{Bur:2005:ASN:}
M.~Bursa, \emph{Astronomische Nachrichten} \textbf{326}, 849--855
  (2005{\natexlab{a}}).

\bibitem[Remilard(2005)]{Rem:2005:ASN:}
R.~A. Remilard, \emph{Astronomische Nachrichten} \textbf{326}, 804--807 (2005).

\bibitem[Strohmayer(2001)]{Str:2001:APJ:}
T.~Strohmayer, \emph{ApJ L} \textbf{552}, L49--L54 (2001),
  \urlprefix\url{http://www.journals.uchicago.edu/ApJ/journal/issues/ApJL/v552%
n1/015001/015001.html}.

\bibitem[Remillard et~al.(2002{\natexlab{a}})]{Rem-etal:2002:APJ:}
R.~A. Remillard, M.~P. Muno., J.~E. McClintock, and J.~Orosz, \emph{ApJ}
  \textbf{580}, 1030--1042 (2002{\natexlab{a}}).

\bibitem[Homan et~al.(2003)]{Hom-etal:2003:ATEL:}
J.~Homan, J.~M. Miller, R.~W. R.~D. Steeghs, T.~Belloni, M.~van~der Klis, and
  W.~H.~G. Lewin (2003), \urlprefix\url{http://integral.rssi.ru/atelmirror/}.

\bibitem[Remillard et~al.(2002{\natexlab{b}})]{Rem-etal:2003:APJ:}
R.~A. Remillard, M.~P. Muno., J.~E. McClintock, and J.~Orosz, \emph{American
  Astronomical Society, HEAD meeting 7} pp. 1030--1042 (2002{\natexlab{b}}).

\bibitem[Aschenbach et~al.(2004)]{Asch-etal:2004:AA:}
B.~Aschenbach, N.~Grosso, D.~Porquet, and P.~Predehl, \emph{A \& A}
  \textbf{417}, 71--78 (2004).

\bibitem[Greene et~al.(2001)]{Gre-etal:2001:APJ:}
J.~Greene, C.~D. Bailyn, and J.~A. Orosz, \emph{ApJ} \textbf{554}, 1290G
  (2001).

\bibitem[Orosz et~al.(2002)]{Oro-etal:2002:APJ:}
J.~Orosz, P.~Groot, M.~van~der Klis, J.~E. McClintock, M.~R. Garcia, P.~Z. P.,
  R.~K. Jain, C.~D. Bailyn, and R.~A. Remillard, \emph{ApJ} \textbf{568}, 8450
  (2002).

\bibitem[Greiner et~al.(2001)]{Gre-etal:2001:NAT:}
J.~Greiner, J.~G. Cuby, and M.~J. McCaughrean, \emph{Nature} \textbf{414}, 522G
  (2001).

\bibitem[McClintock and Remillard(2004)]{McCRem:2004:CXS:}
J.~E. McClintock, and R.~A. Remillard, \enquote{Black Hole Binaries,} in
  \emph{Compact Stellar X-Ray Sources}, edited by W.~H.~G. Lewin, and
  M.~van~der Klis, Cambridge University Press, Cambridge, 2004, e-print 2003,
  \eprint{astro-ph/0306213}.

\bibitem[Beer and Podsiadlowski(2002)]{BeePod:2002:MNRAS:}
M.~E. Beer, and P.~Podsiadlowski, \emph{MNRAS} \textbf{331}, 351 (2002).

\bibitem[Abramowicz et~al.(2004)]{Abr-etal:2004:APJ:}
M.~Abramowicz, W.~Klu{\'z}niak, J.~E. McClintock, and R.~A. Remillard,
  \emph{ApJ L} \textbf{609}, L63--L65 (2004).

\bibitem[T\"or\"ok(2005{\natexlab{a}})]{Tor:2005:AA:}
G.~T\"or\"ok, \emph{A\&A} \textbf{440}, 1--4 (2005{\natexlab{a}}).

\bibitem[Nowak and Lehr(1998)]{NowLeh:1998:TAD:}
M.~A. Nowak, and D.~E. Lehr, \enquote{{S}table oscillations of black hole
  accretion discs,} in \emph{{T}heory of {B}lack {H}ole {A}ccretion {D}isks},
  edited by M.~A. Abramowicz, G.~Bj{\"{o}}rnsson, and J.~E. Pringle, Cambridge
  University Press, Cambridge, 1998, pp. 233--253, \eprint{astro-ph/9812004}.

\bibitem[Abramowicz and Klu\'zniak(2001)]{AbrKlu:2001:AA:}
M.~A. Abramowicz, and W.~Klu\'zniak, \emph{A\&A L} \textbf{374L}, 19A (2001).

\bibitem[T\"or\"ok et~al.(2005)]{Tor-etal:2005:AA:}
G.~T\"or\"ok, M.~A. Abramowicz, W.~W. Klu{\'z}niak, and Z.~Stuchl{\'{\i}}k,
  \emph{A\&A} \textbf{436}, 1--8 (2005), \eprint{astro-ph/0401464}.

\bibitem[T\"or\"ok(2005{\natexlab{b}})]{Tor:2005:ASN:}
G.~T\"or\"ok, \emph{Astronomische Nachrichten} \textbf{326}, 856--860
  (2005{\natexlab{b}}).

\bibitem[Shafee et~al.(2006)]{Sha-etal:2006:APJ:}
R.~Shafee, J.~E. McClintock, R.~Narayan, S.~W. Davis, Li, and R.~A. Remillard,
  \emph{APJ L} \textbf{636}, L113=L116 (2006).

\bibitem[Bursa(2005{\natexlab{b}})]{Bur:2005:RAG:}
M.~Bursa, \enquote{,} in \emph{Proceedings of RAGtime 6/7: Workshops on black
  holes and neutron stars, Opava, 16--18/18--20 September 2004/2005}, edited by
  S.~Hled\'{\i}k, and Z.~Stuchl\'{\i}k, Silesian University in Opava, Opava,
  2005{\natexlab{b}}, ISBN 80-7248-242-4.

\bibitem[Abramowicz et~al.(2003{\natexlab{a}})]{Abr-etal:2003:AA:}
M.~A. Abramowicz, T.~Bulik, M.~Bursa, and W.~Klu{\'z}niak, \emph{A\&A L}
  \textbf{404}, L21--L24 (2003{\natexlab{a}}).

\bibitem[Psaltis et~al.(1998)]{Psa-etal:1998:ASTRJ2L:}
D.~Psaltis, M.~Mendez, R.~Wijnands, J.~Homan, P.~G. Jonker, M.~van~der Klis,
  F.~K. Lamb, E.~Kuulkers, J.~van Paradijs, and W.~H.~G. Lewin, \emph{ApJ L}
  \textbf{501}, L95 (1998), \eprint{astro-ph/9805084}.

\bibitem[Abramowicz et~al.(2003{\natexlab{b}})]{Abr-etal:2003:PUBASJ:}
M.~A. Abramowicz, V.~Karas, W.~Klu{\'z}niak, W.~Lee, and P.~Rebusco,
  \emph{PASJ} \textbf{55}, 467--471 (2003{\natexlab{b}}),
  \urlprefix\url{http://pasj.asj.or.jp/v55/n2/550214/550214-frame.html}.

\bibitem[Rebusco(2004)]{Reb:2004:PUBASJ:}
P.~Rebusco, \emph{PASJ} \textbf{56}, 553--557 (2004).

\bibitem[Hor\'{a}k(2004)]{Hor:2004:RAGtime4and5:Proceedings:}
J.~Hor\'{a}k, \enquote{General apects of nonlinear resonance $3\!:\!2$ in QPO
  context,} in \emph{Proceedings of RAGtime 4/5: Workshops on black holes and
  neutron stars, Opava, 14--16/13--15 October 2002/03}, edited by
  S.~Hled{\'{\i}}k, and Z.~Stuchl{\'{\i}}k, Silesian University in Opava,
  Opava, 2004.

\bibitem[Belloni et~al.(2005)]{Bel-Men-Hom:2005:ASTRA:}
T.~Belloni, M.~M{\'e}ndez, and J.~Homan, \emph{A\&A} \textbf{437}, 209--216
  (2005).

\bibitem[Abramowicz et~al.(2005)]{Abr-etal:2005:RAGtime6and7:}
M.~A. Abramowicz, D.~Barret, M.~Bursa, J.~Hor{\'a}k, W.~Klu{\'z}niak,
  P.~Rebusco, and G.~T\"or\"ok, \enquote{,} in \emph{Proceedings of RAGtime
  6/7: Workshops on black holes and neutron stars, Opava, 16--18/18--20
  September 2004/2005}, edited by S.~Hled\'{\i}k, and Z.~Stuchl\'{\i}k,
  Silesian University in Opava, Opava, 2005, ISBN 80-7248-242-4.

\bibitem[Abramowicz et~al.(2006)]{Abr-etal:2006:MNRAS:}
M.~A. Abramowicz, D.~Barret, M.~Bursa, J.~Hor{\'a}k, W.~Klu{\'z}niak, J.~F.
  Olive, P.~Rebusco, and G.~T\"or\"ok  (2006), in preparation, to be submitted
  to MNRAS.

\bibitem[Boirin et~al.(2000)]{Boi-etal:2000:}
L.~Boirin, D.~Barret, J.~G. Olive, P.~F. Bloser, and J.~E. Grindlay, Low and
  high frequency quasi-periodic oscillations in 4u1915-05 (2000),
  \eprint{astro-ph/0007071}.

\bibitem[Homan et~al.(2002)]{Hom-etal:2002:ASTRJ2:}
J.~Homan, M.~van~der Klis, P.~G. Jonker, R.~Wijnands, E.~Kuulkers,
  M.~M{\'e}ndez, and W.~H.~G. Lewin, \emph{ApJ} \textbf{568}, 878--900 (2002).

\bibitem[Jonker et~al.(2000)]{Jon-etal:2000:ASTRJ2:}
P.~G. Jonker, M.~van~der Klis, R.~Wijnands, J.~van Paradijs, M.~M{\'e}ndez,
  E.~C. Ford, E.~Kuulkers, and F.~K. Lamb, \emph{ApJ} \textbf{537}, 374--386
  (2000).

\bibitem[Jonker et~al.(2002)]{Jon-etal:2002:MONNR:}
P.~G. Jonker, M.~van~der Klis, J.~Homan, M.~M{\'e}ndez, W.~H.~G. Lewin,
  R.~Wijnands, and W.~Zhang, \emph{MNRAS} \textbf{333}, 665 (2002).

\bibitem[T\"or\"ok et~al.(2006)]{Tor-etal:2006:INPR:}
G.~T\"or\"ok, D.~Barret, and J.~Hor{\'a}k  (2006), in preparation.

\bibitem[Hor{\'a}k et~al.(2006)]{Hor-etal:2006:INPR:}
J.~Hor{\'a}k, M.~A. Abramowicz, and G.~T\"or\"ok  (2006), in preparation.

\bibitem[Barret et~al.(2005{\natexlab{a}})]{Bar-etal:2005:AN:}
D.~Barret, J.~F. Olive, and M.~C. Miller, \emph{Astronomische Nachrichten}
  \textbf{326}, 808--811 (2005{\natexlab{a}}).

\bibitem[Barret et~al.(2005{\natexlab{b}})]{Bar-etal:2005:MNRAS:}
D.~Barret, J.~F. Olive, and M.~C. Miller, \emph{MNRAS} \textbf{361}, 855--860
  (2005{\natexlab{b}}).

\end{thebibliography}
\end{document}